\documentclass[aps,twocolumn,preprintnumbers,amsart]{revtex4}

\usepackage{color}
\usepackage{amssymb}
\usepackage{wasysym}
\usepackage{graphicx}
\usepackage{appendix}
\usepackage{natbib}
\usepackage{hyperref}
\usepackage[tight,FIGBOTCAP]{subfigure}

\begin{document}
\title{Efficient long distance quantum communication}
 \author{Sreraman Muralidharan$^{1*}$}
 \author{Linshu Li$^{2*}$}
\author{Jungsang Kim$^{3}$}
\author{Norbert L\"utkenhaus$^{4}$}
\author{Mikhail D. Lukin$^{5}$}
\author{Liang Jiang$^{2}$}
\affiliation{$^1$Department of Electrical Engineering, Yale University, New Haven, CT
06511 USA}
\affiliation{$^2$Department of Applied Physics, Yale University, New Haven, CT 06511 USA}
\affiliation{$^3$Department of Electrical and Computer Engineering, Duke University,
Durham, NC 27708 USA}
\affiliation{$^4$Institute of Quantum computing, University of Waterloo, N2L 3G1
Waterloo, Canada}
\affiliation{$^5$Department of Physics, Harvard University, Cambridge, MA 02138, USA}
\date{\today }
\thanks{Equal contribution}
\pacs{03.67.Dd, 03.67.Hk, 03.67.Pp.}
\begin{abstract}
Despite the tremendous progress of quantum cryptography, 
efficient quantum communication over long distances ($\ge1000$km) remains an outstanding challenge
due to fiber attenuation and operation errors accumulated over the
entire communication distance. Quantum repeaters (QRs), as a promising approach,
can overcome both photon loss and operation errors, and hence significantly speedup
the communication rate. Depending on the methods used to correct loss
and operation errors, all the proposed QR schemes can be classified into three categories (generations). Here we present the first systematic
comparison of three generations of quantum repeaters by evaluating
the cost of both temporal and physical resources, and identify the
optimized quantum repeater architecture for a given set of experimental
parameters. Our work provides a roadmap for the experimental realizations of highly efficient quantum
networks over transcontinental distances. 
\end{abstract}
\maketitle
\section{Introduction}
First developed in the 1970s, fiber-optic communication systems have
boosted the rate of classical information transfer and played a major
role in the advent of the information age. The possibility to encode
information in quantum states using single photons and transmit them
through optical channels has led to the
development of quantum key distribution (QKD) systems \cite{Lo14}. However,
errors induced by the intrinsic channel attenuation, i.e. loss errors, become a major barrier for efficient quantum
communication over continental scales, due to the exponential decay of communication rate
 \cite{Takeoka2014}. In contrast to classical communication, due to the quantum no-cloning theorem \cite{Wootters82}, 
quantum states of photons cannot be amplified without any disturbance. In addition to loss errors, depolarization errors introduced by the imperfect optical channel can impair the quality of the single photon transmitted and hence the quantum information encoded.

To overcome these challenges, quantum
repeaters (QRs) have been proposed for the faithful realization of long-distance quantum communication \cite{Briegel1998}.
The essence of QRs is to divide the total distance of communication
into shorter intermediate segments connected by QR stations, in which loss
errors from fiber attenuation can be corrected. Active mechanisms 
are also employed at every repeater station to correct operation errors, i.e. imperfections induced by the channel, measurements and gate operations.

\begin{figure*}
\centering \includegraphics[width=15cm]{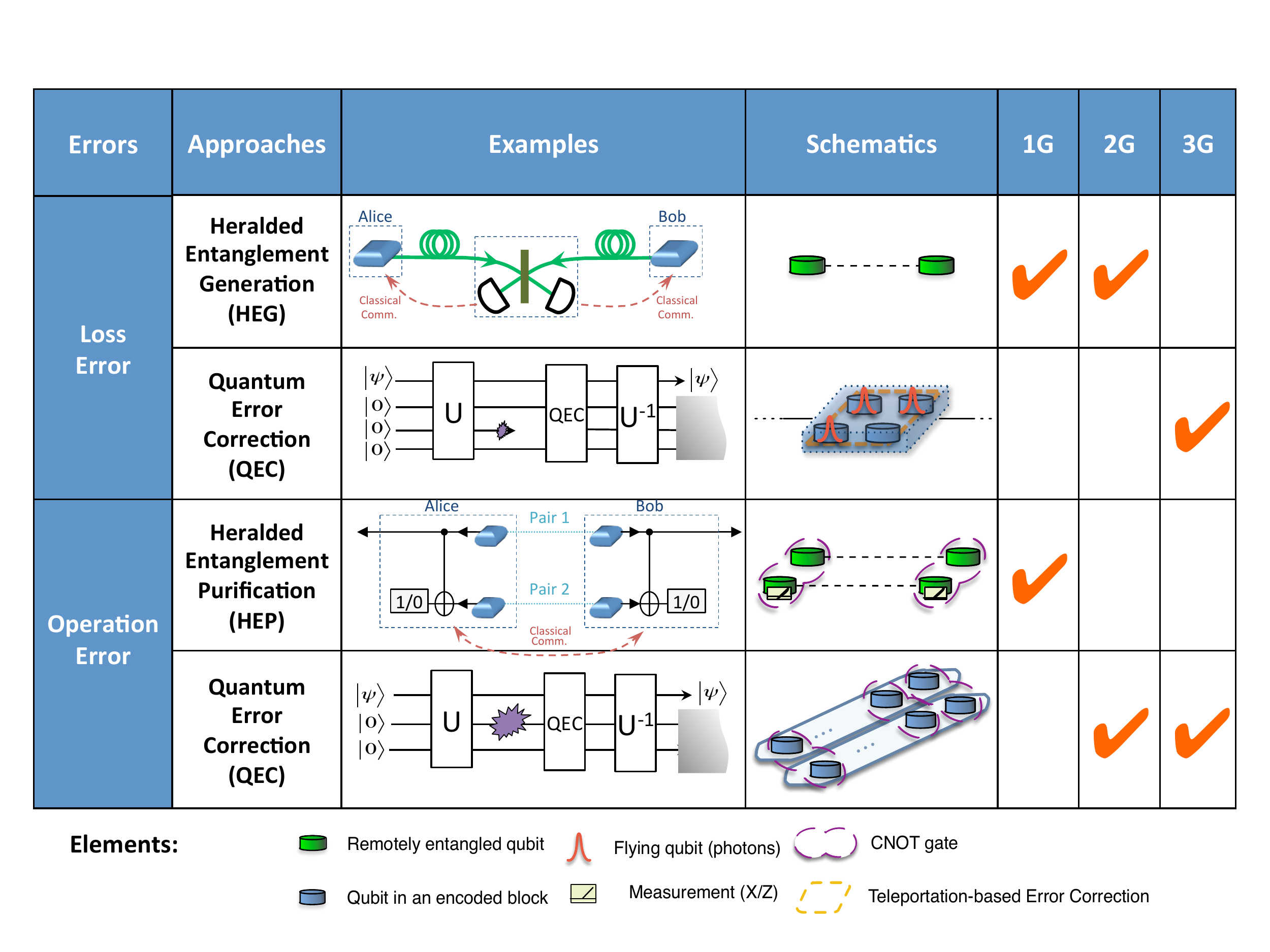}
\protect\caption{A list of methods to correct loss and operation errors. Depending
on the methods used to correct the errors, QRs are categorized into three generations.}
\label{fig:generations2} 
\end{figure*}

As illustrated in Fig.~1, loss errors can be suppressed by either heralded entanglement
generation (HEG) \cite{Briegel1998,Sangouard2011} or quantum error correction
(QEC) \cite{JTNMVL09,Munro10,Fowler2010,Muralidharan2014,Muralidharan2015}. During HEG, quantum entanglement can be generated with techniques such as two-photon interference conditioned on the click patterns of the detectors in between. 
Loss errors are suppressed by repeating this heralded
proceDure until the two adjacent stations receive the confirmation of certain
successful detection patterns via \textit{two-way} classical signaling.

Alternatively, one may encode the logical qubit into a block of physical
qubits that are sent through the lossy channel and use quantum error correction to restore the logical qubit with only \textit{one-way} signaling. 
Quantum error correcting codes can correct no more than $50\%$ loss rates deterministically due to the no-cloning theorem  \cite{Stace09,Muralidharan2014}. 
To suppress operation errors, one may use either heralded entanglement purification (HEP) \cite{Deutsch96,Dur99} or QEC \cite{JTNMVL09,Munro10,Fowler2010,Muralidharan2014,Muralidharan2015} as listed in Fig. \ref{fig:generations2}. In HEP, multiple low-fidelity Bell pairs are consumed to probabilistically generate a smaller number of higher-fidelity Bell pairs. Like HEG, to confirm the success of purification , \textit{two-way} classical signaling between repeater stations for exchanging measurement results is required.  Alternatively, QEC can correct operation errors using only \textit{one-way} classical
signaling, but it needs high fidelity local quantum gates.

\begin{figure*}
\centering \includegraphics[width=17cm]{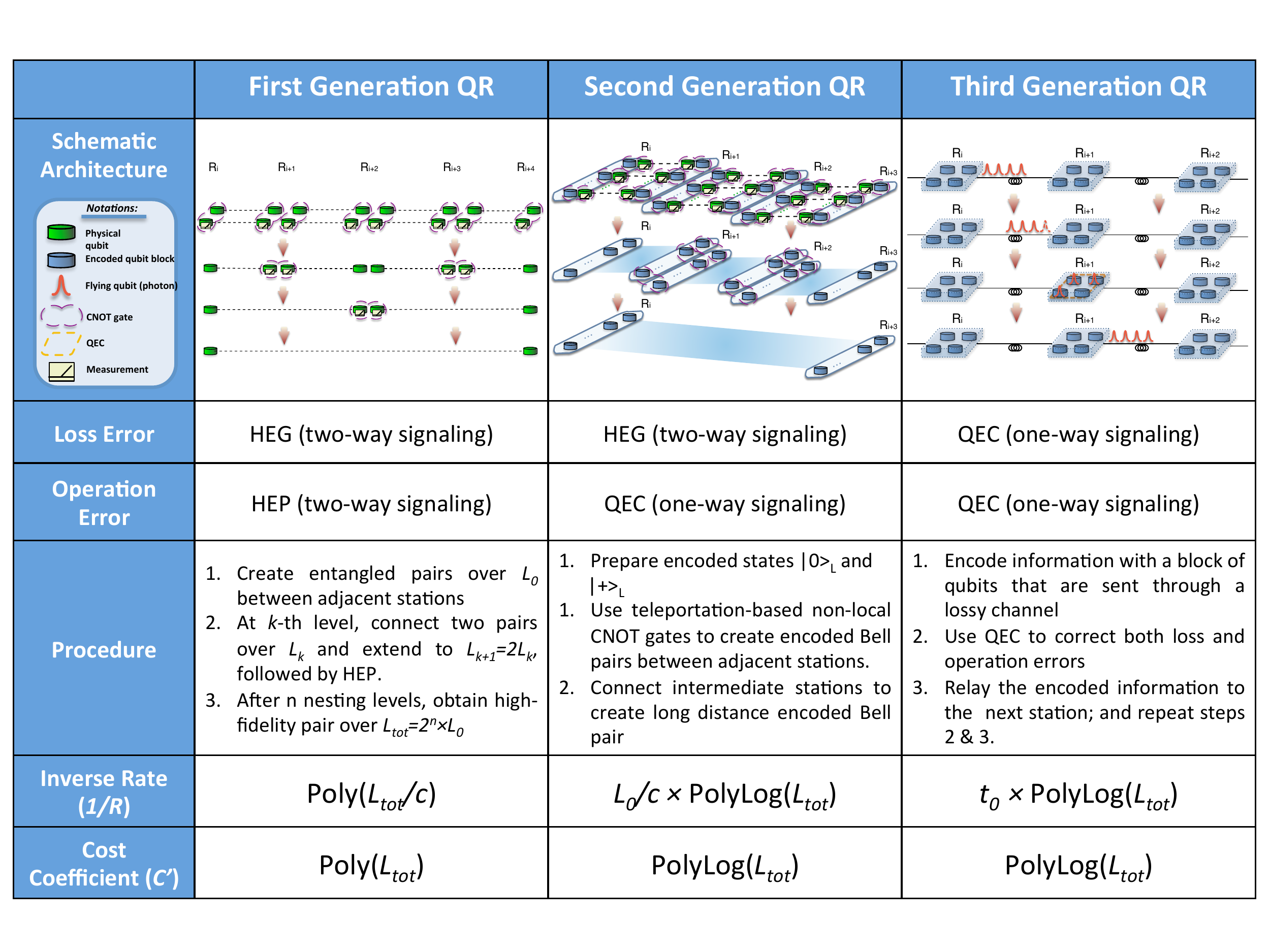}
\protect\caption[fig:generations]{Comparison of three generations of QRs.}
\label{fig:generations} 
\end{figure*}

Based on the methods adopted to suppress loss and operation errors,
we can classify various QRs into three 
categories as shown schematically in Fig.~\ref{fig:generations}, which we refer to as three generations of QRs \cite{Munro2015} \footnote{Note that the combination of QEC for loss errors and HEG for operation errors is sub-optimum compared to the other three combinations.}.
Each generation of QR performs the best for a specific regime of operational parameters such as local gate speed, gate fidelity, and coupling efficiency. In this paper, we consider both the temporal and physical resources consumed by the three generations of QRs and identify the most efficient architecture for different parameter regimes. The results can guide the design of efficient long distance quantum communication links that act as elementary building blocks for future quantum networks.

The paper is organized as follows: In the following section, we will briefly review the chacracterstics of three generations of QRs. In section III, we use the cost coefficient as an optimization metric to compare the QR performance, and study its dependence on the individual operational parameters including coupling efficiencies, gate fidelities, and gate speed. In section IV, we present a holistic view of the optimization and illustrate the parameter regions where each generation of QRs perform more efficiently than others. In section V, we analyze the advantages and challenges of each generation of QRs and discuss the experimental candidates for their realizations.
\normalcolor
\section{Three generations of quantum repeaters }
The first
generation of QRs uses HEG and HEP to suppress loss and operation
errors, respectively \cite{Briegel1998,Sangouard2011}. This approach starts with
purified high-fidelity entangled pairs with separation $L_{0}=L_{tot}/2^{n}$
created and stored in adjacent stations. At $k$-th nesting level,
two entangled pairs of distance $L_{k-1}=2^{k-1}L_{0}$ are connected
to extend entanglement to distance $L_{k}=2^{k}L_{0}$
 \cite{Zukowski93}. As practical gate operations and entanglement swapping inevitably cause the fidelity of entangled pairs to drop, HEP can be incorporated at each level of entanglement extension \cite{Deutsch96,Dur99}. With $n$ nesting levels of connection
and purification, a high-fidelity entangled pair over distance $L_{n}=L_{tot}$
can be obtained. The first generation of QRs reduces the exponential overhead in direct state transfer to only polynomial overhead, which is limited by the two-way classical signaling required by HEP between non-adjacent repeater stations. The communication rate still decreases polynomially with distance and thus becomes very slow for long distance quantum communication. The communication rate of the
first generation of QRs can be boosted using temporal, spatial, and/or frequency multiplexing associated with the internal degrees of {freedom for the} quantum memory \cite{Sangouard2011,Bonarota11}.

The second generation of QRs uses HEG to suppress loss errors and
QEC to correct operation errors \cite{JTNMVL09,Munro10, Brinew2}. First,
the encoded states $|0\rangle_{L}$ and $|+\rangle_{L}$ are fault-tolerantly 
prepared using the Calderbank-Shor-Steane (CSS) codes \footnote{CSS codes are considered because of the
fault tolerant implementation of preparation, measurement, and
encoded CNOT gate \cite{JTNMVL09,NC00}} and stored at two adjacent stations. Then, an encoded Bell
pair $|\Phi^{+}\rangle_{L}=\frac{1}{\sqrt{2}}\left(|0,0\rangle_{L}+|1,1\rangle_{L}\right)$
between adjacent stations can be created using teleportation-based
non-local CNOT gates \cite{Gottesman99,JTSL07b} applied to each physical qubit in the encoded block using the entangled pairs generated through HEG process. 
Finally, QEC is carried out when entanglement swapping at the encoded level is performed to extend the range of entanglement. The second generation uses QEC to replace HEP and therefore avoids the time-consuming two-way classical
signaling between non-adjacent stations. The communication rate is then limited by the time delay associated with two-way classical signaling between adjacent stations and local gate operations. 
If the probability of accumulated operation errors over all repeater stations is sufficiently small, we can
simply use the second generation of QRs \textit{without} encoding.

The third generation of QRs relies on QEC to correct both loss and
operation errors \cite{Fowler2010,Munro2012a,Muralidharan2014,Muralidharan2015}. The quantum
information can be directly encoded in a block of physical qubits
that are sent through the lossy channel. If the loss and operation errors are sufficiently small, the received physical qubits can be used to restore the whole encoding block, which is retransmitted to the next repeater station. The third generation of QRs only needs \textit{one-way} signaling and thus
can achieve very high communication rate, just like the classical
repeaters only limited by local operation delay. It turns out that
quantum parity codes \cite{Ralph05} with moderate coding blocks (\textasciitilde{}200
qubits) can efficiently overcome both loss and operation errors \cite{Muralidharan2014,Munro2012a}. 

Note that the second and third generations of QRs can achieve communication
rate much faster than the first generation over long distances, but
they are technologically more demanding. For example, they require
high fidelity quantum gates as QEC only works
well when operation errors are below the fault tolerance threshold. The repeater spacing for the third generation of QRs is smaller compared to the first two generation of QRs because error correction can only correct a finite amount of loss errors. 
Moreover, quantum
error correcting codes can correct only up to 50$\%$ loss error rates deterministically,
which restricts the applicable parameter range for the third generation
of QRs \cite{Muralidharan2014}.

\section{Comparison of three generations of QRs}

To present a systematic comparison of different  in terms of efficiency, we need to consider
both temporal and physical resources. The temporal resource depends on the rate, which is limited by the time delay from the two-way classical signaling (first and second generations) and the local gate operation (second and third generations) \cite{JTKL07}. The physical
resource depends on the total number of qubits needed for HEP (first and second generations) and QEC
(second and third generations) \cite{Muralidharan2014,Bratzik14}. We
propose to quantitatively compare the three generations of QRs
using a cost function  \cite{Muralidharan2014} related to the required number of qubit memories to achieve a given transmission rate. 
Suppose a total
of $N_{tot}$ qubits are needed to generate secure keys at $R$ bits/second, the cost function is defined as 
\begin{equation}
C(L_{tot})=\frac{N_{tot}}{R}=\frac{N_{s}}{R}\times \frac{L_{tot}}{L_{0}} ,
\end{equation}
where $N_{s}$ is the number of qubits needed per repeater station,
$L_{tot}$ the total communication distance, and $L_{0}$ the
spacing between neighboring stations. Since the cost function scales at least linearly
with $L_{tot}$, to demonstrate the additional overhead associated with $L_{tot}$, the \textit{cost coefficient} can be introduced as
\begin{equation}
C'(L_{tot})=C/L_{tot},
\end{equation}
which can be interpreted as the resource overhead (qubits
$\times$ time) for the creation of one secret bit over 1km (with
target distance $L_{tot}$). Besides the fiber attenuation (with $L_{att}=20$
km), the cost coefficient also depends on other experimental parameters,
in particular the coupling efficiency $\eta_{c}$ (see supplementary material), the
gate error probability $\epsilon_{G}$, and the gate time $t_{0}$ \footnote{For simplicity, we assume that the fidelity of physical Bell pairs $F_{0}=1-\frac{5}{4}\epsilon_{g}$
achieved with entanglement purification and measurement error probability $\epsilon_m = \frac{\epsilon_g}{4}$ through a verification proceDure.}. For third generation QRs, we restrict the search only up to 200 qubits per logical qubit considering the complexity involved in the production of larger codes and for a fair comparison with second generation of QRs. For simplicity, we will assume that $t_0$ is independent of code size for small encoded blocks for second and third generation QRs.
We will now investigate how $C'$ varies with these parameters
for three generations of QRs, and identify the optimum generation
of QR depending on the technological capability.

\subsection{Coupling efficiency}
\begin{figure}
\centering \includegraphics[width=8cm,height=11cm]{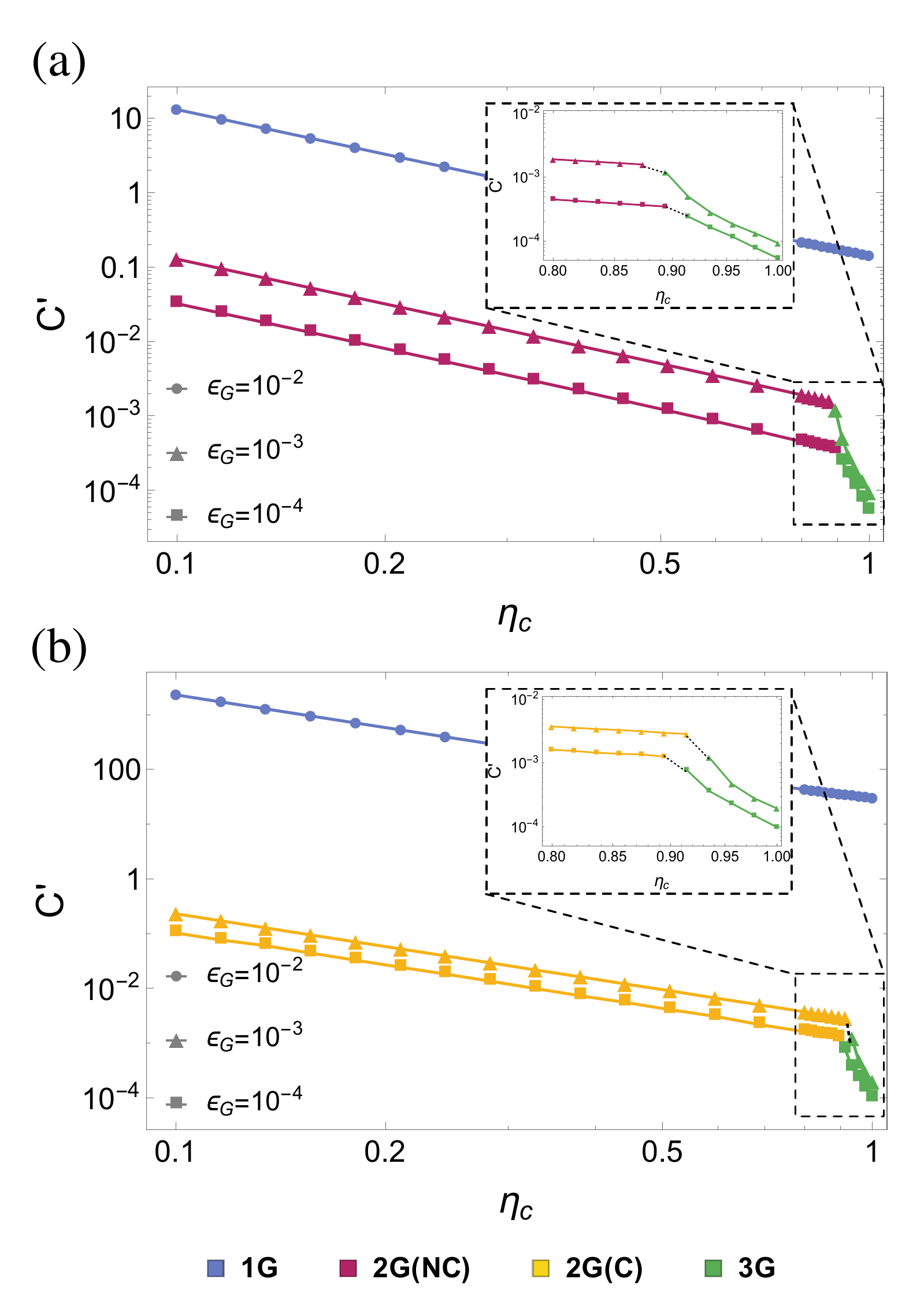}
\protect\caption{The optimized cost coefficient C' as a function of $\eta_{c}$ for
$t_{0}=1\mu s$, $\epsilon_{G}\in\left\{ 10^{-2},10^{-3},10^{-4}\right\} $,
and a) $L_{tot}=1000$km, b) $L_{tot}=10,000$km. The associated optimized
QR protocols are indicated in different colors.}

\label{fig:etac} 
\end{figure}

The coupling efficiency $\eta_{c}$ accounts for 
the emission of photon from the memory qubit, coupling of the photon into the optical fiber and vice versa, and the final detection of photons. The first
and second generations of QRs use HEG compatible with arbitrary coupling
efficiency, while the third generation relies on QEC requiring the overall transmission (including the 
coupling efficiency $\eta_{c}$ and the channel transmission)  to be at least above 50\% \cite{Bennett97,Muralidharan2014}. 
As illustrated in Fig.~\ref{fig:etac}, for high coupling efficiency
($\eta_{c}\apprge90\%$) the third generation of QRs has an obvious advantage
over the other generations due to the elimination of two-way classical
signaling. As the coupling efficiency is reduced and approaches ($\sim90\%$) for quantum parity codes, the size of the coding block quickly increases and it becomes less favorable to use this approachsince we restrict the size of the encoded block for third generation of QRs. For coupling efficiency below $\sim90\%$,
the optimization chooses the first and second generations of QRs, and
then $C'$ is proportional to $\eta_{c}^{-2}$ for HEG
protocols heralded by two-photon detector click patterns.

If the gate error becomes large (e.g., $\epsilon_{G}=10^{-2}$),
the capability of correcting loss errors will be compromised for the
third generation QRs. Similar trends can be observed
as we fix $\epsilon_{G}$ and increase $L_{tot}$. In contrast to
the third generation QRs, the first and second generation QRs with HEG works well even for low coupling efficiencies.

\subsection{Speed of quantum gates}

\begin{figure}
\centering \includegraphics[width=8cm,height=11cm]{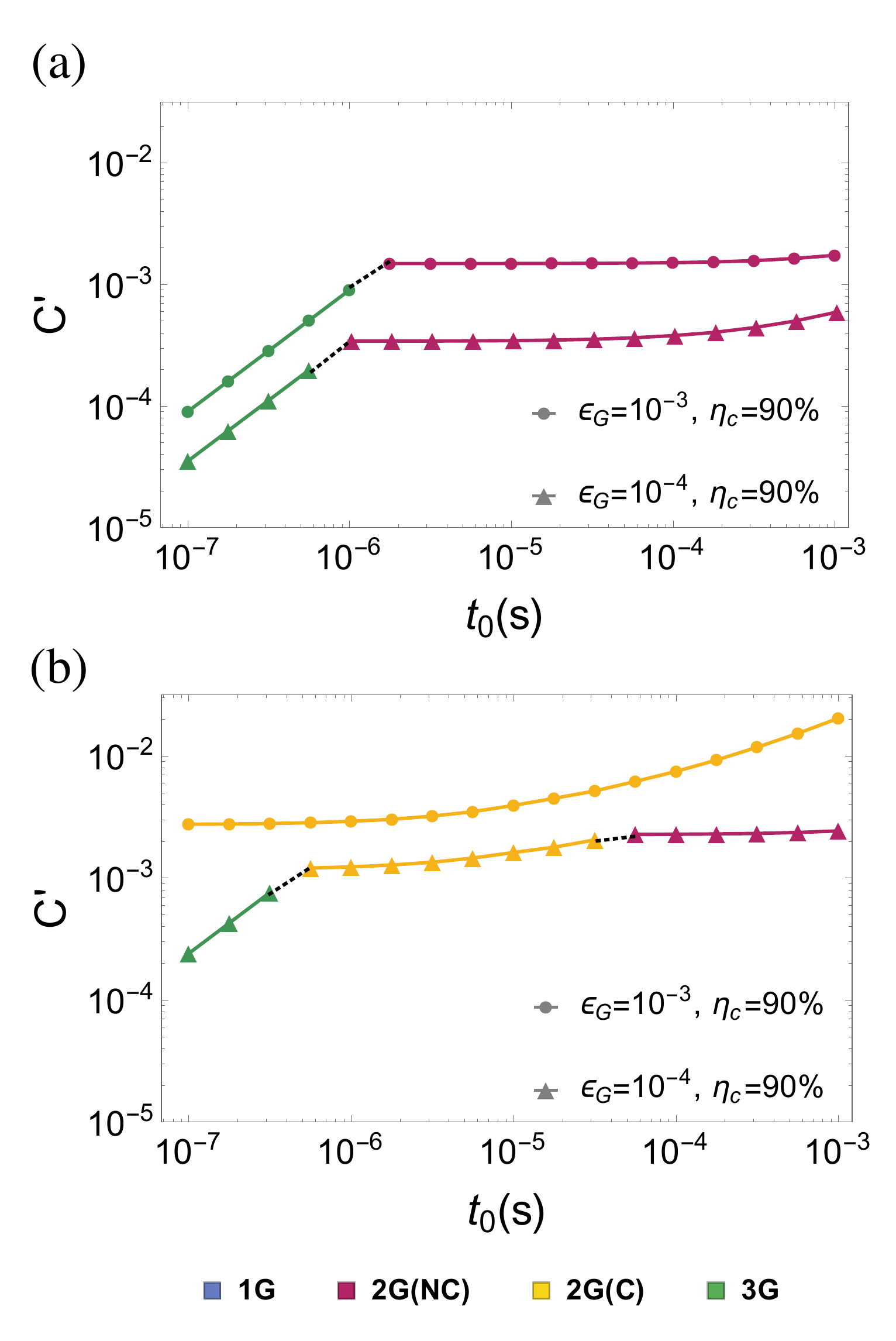}
\protect\caption{The optimized cost coefficient C' as a function of $t_{0}$ for $\eta_{c}=0.9$,
$\epsilon_{G}\in\left\{ 10^{-3},10^{-4}\right\} $, and a) $L_{tot.}=1000$km,
b) $L_{tot.}=10,000$km. The associated optimized QR protocols are
indicated in different colors.}
\label{fig:speed} 
\end{figure}

We investigate the performance of different generations of QRs for different
gate times in the range $0.1\mu s\leq t_{0} \leq 100\mu s$. As shown in Fig.~\ref{fig:speed}, for high speed quantum
gates ($t_{0}\apprle 1\mu s$) the third generation of QRs provides
a very fast communication rate, which makes it the most favorable
protocol, with $C'\propto t_{0}$. For slower quantum gates ($t_{0}\apprge10\mu s$),
the gate time becomes comparable or even larger than the delay of
two-way classical signaling between adjacent stations ($t_{0}\apprge\frac{L_{0}}{c}\approx\frac{L_{att}}{c}$);
as the third generation of QRs loses its advantage in communication rate,
the second generation of QRs with less physical resources becomes the optimized
QR protocol, with almost constant $C'$ for a wide range of $t_{0}$.

We notice that for small gate error and intermediate distance (e.g.,
$\epsilon_{G}=10^{-4}$ and $L_{tot}=1000$km appeared in Fig.~\ref{fig:etac}a
and \ref{fig:speed}a), encoding might not even be necessary for the
second generation of QRs, because the accumulated errors over the entire
repeater network are within the tolerable range for quantum communication
($\epsilon_{G}\frac{L_{tot}}{L_{att}}\apprle0.1$). However, for larger
error probability or longer distances ($\epsilon_{G}\frac{L_{tot}}{L_{att}}\gg0.1$),
encoding is required for the second generation QRs\footnote{When $\epsilon_{G}$ increases from $10^{-4}$ to $10^{-3}$, the cost coefficient for the second generation of QR without encoding
increases by almost a factor of 10 (Fig.~\ref{fig:speed}a), while
the change is less significant for the second and third generations
of QRs with encoding (Fig.~\ref{fig:speed}). This is because at
the logical level, the change in the effective logical error probability
is suppressed for the given set of parameters.}. The cost coefficient for the first generation of QRs ($C'>1\frac{qubit\times sec}{sbit\times km}$)
lies beyond the scope of Fig.~\ref{fig:speed}, with little dependence
on $t_{0}$ that is mostly negligible compared to the two-way classical
signaling between non-adjacent stations ($\frac{L_{tot}}{c}>10ms$).

\subsection{Gate fidelity}

\begin{figure}
\centering \includegraphics[width=8cm,height=11cm]{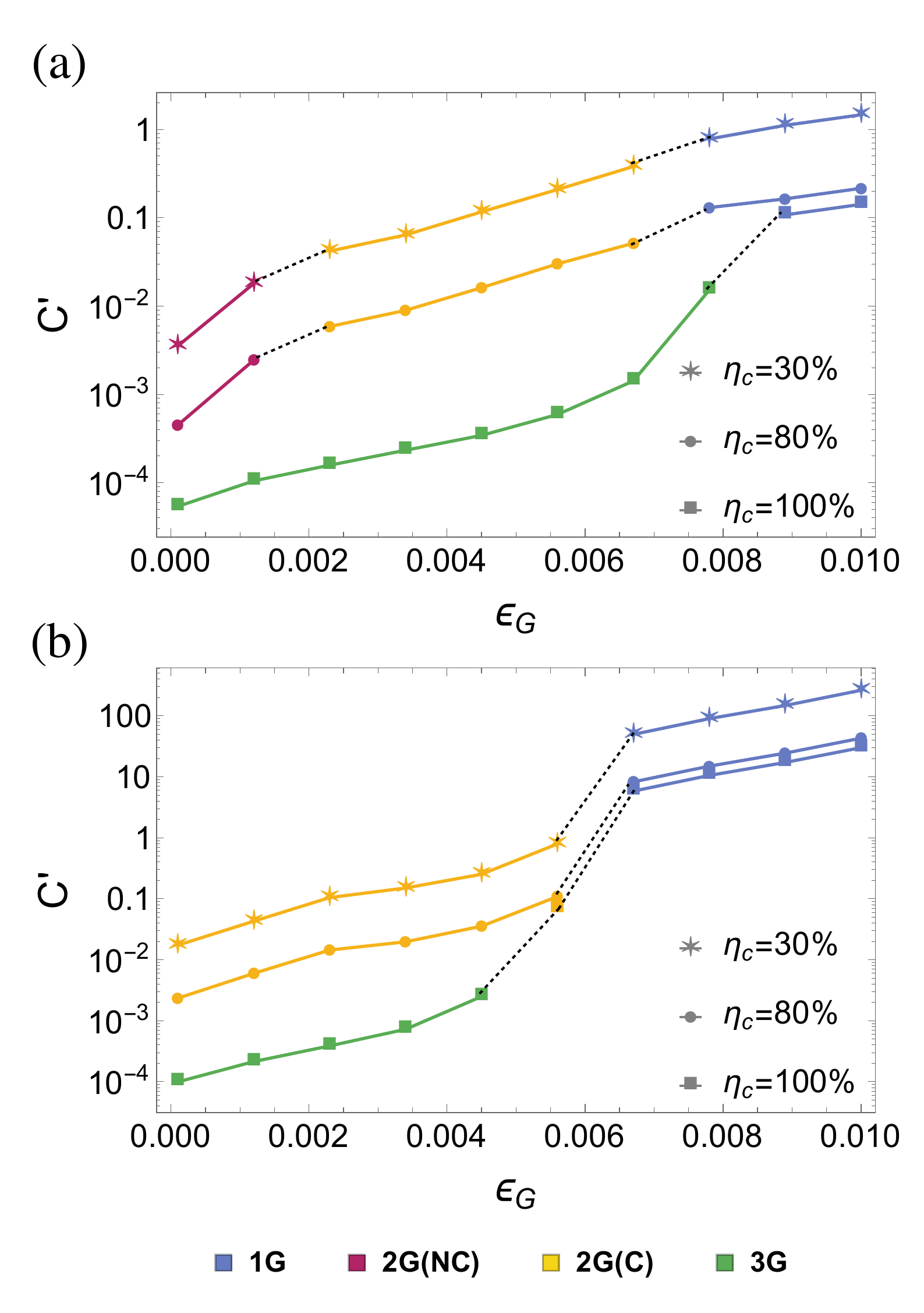}
\protect\caption{The optimized cost coefficient C' as a function of $\epsilon_{G}$
for $t_{0}=1\mu s$, $\eta_{c}\in\left\{ 30\%,80\%,100\%\right\} $,
and a) $L_{tot}=1000$km, b) $L_{tot}=10,000$km. The associated optimized
QR protocols are indicated in different colors.}
\label{fig:gate} 
\end{figure}

The three generations of QRs have different thresholds in terms of
gate error probability $\epsilon_{G}$. The first generation relies
on HEP with the highest operation error threshold
up to about $3\%$ \cite{Briegel1998}. The second and third generations
both use QEC to correct operation errors, with error correction thresholds
of approximately $1\%$ \cite{Knill05}. The gate error threshold
of the second generation is slightly lower than that of the third
generation, because of the extra gates required for teleportation-based
non-local CNOT gates and entanglement swapping in the second generation of QRs (See supplementary material). However, since we restrict the size of the encoded block for third generation of QRs, C' increases exponentially with $\epsilon_G$ slightly below the theoretical threshold of quantum parity codes.
As illustrated in Fig.~\ref{fig:gate}, for almost perfect coupling efficiency
(e.g. $\eta_{c}=100\%$) and fast local operation ($t_{0}=1\mu s$),
the third generation using QEC to correct both fiber attenuation loss and operation
errors is the optimized protocol for moderate gate errors.
For lower coupling efficiencies (e.g. $\eta_{c}=30\%$ and $80\%$)
with too many loss errors for the third generation to tolerate, the
first and second generations with HEG yield
good performance. As $\epsilon_{G}$ increases, there is a transition at
about $0.8\%$($0.6\%$) below which the second generation is more favorable for 1000km (10000km).

\section{Optimum generation of QRs}

\begin{figure*}
\centering \includegraphics[width=19cm,height=14cm]{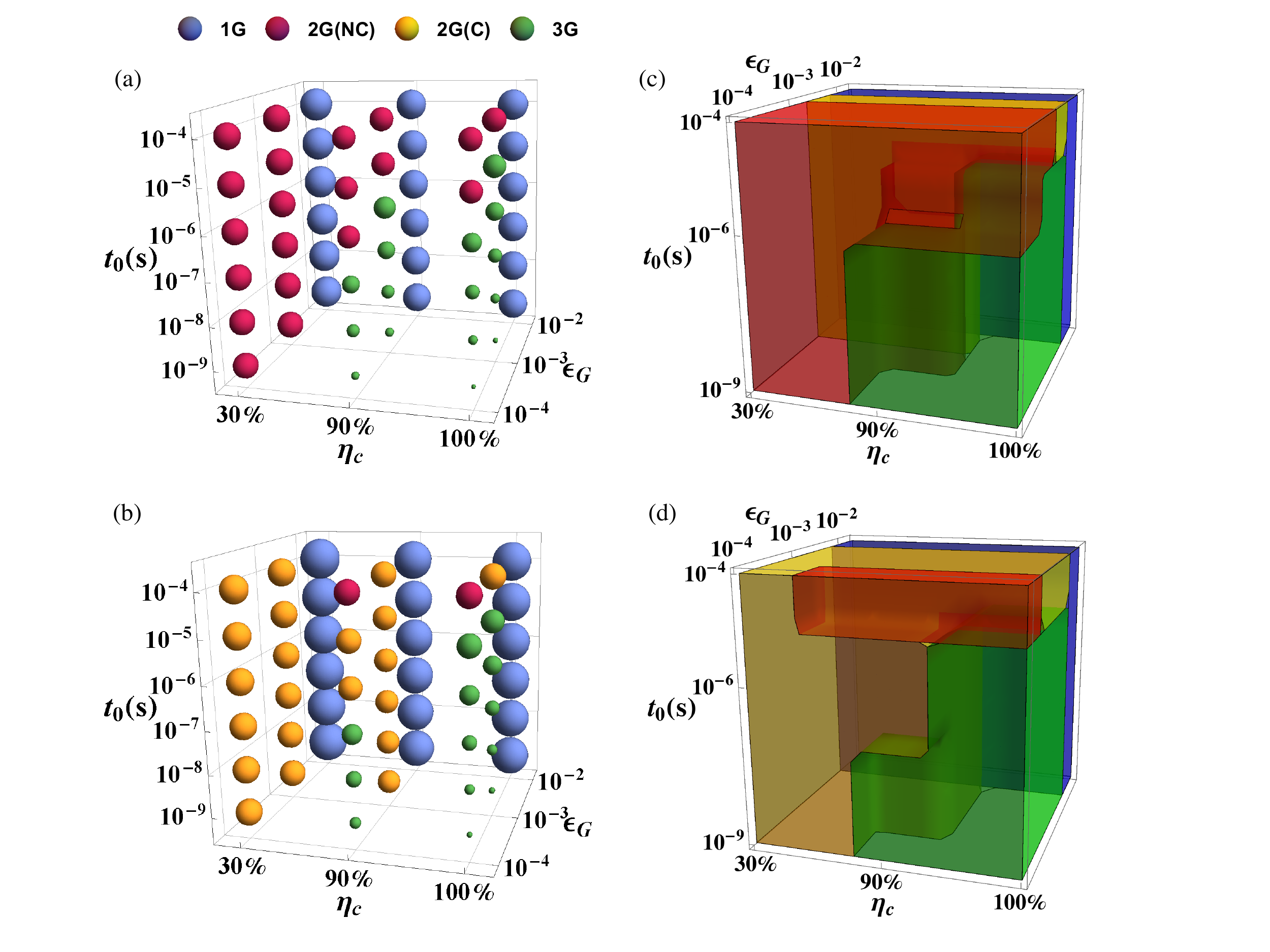}
\protect\caption{ The bubble plot comparing various QR protocols in the three-dimensional parameter
space spanned by $\eta_{c}$, $\epsilon_{G}$, and $t_{0}$, for a)
$L_{tot}=1000$km and b) $L_{tot}=10,000$km. The bubble color indicates
the associated optimized QR protocol, and the bubble diameter is proportional
to the cost coefficient. The region plots (c) and (d) showing the distribution of different optimized QR protocol in the three dimensional parameter space for $L_{tot}=1000$km and $L_{tot}=10,000$km respectively. The region plot (c) contains a yellow region of second generation with encoding, which can be verified in a bubble plot with a finer discretization of $\epsilon_G$.
}
\label{fig:bubble} 
\end{figure*}

Based on the above analysis of the cost coefficient that depends on
the coupling efficiency $\eta_{c}$, the gate time $t_{0}$, and the
gate infidelities $\epsilon_{G}$, we may summarize the results using
the bubble plot and the region plot in the three-dimensional parameter space, as shown
in Fig. \ref{fig:bubble}. The bubble color indicates the associated
optimized QR protocol, and the bubble diameter is proportional to the
cost coefficient. The parameter space can be divided into the following
regions: (I) For high gate error probability $(\epsilon_{G}\gtrsim 1\%)$,
the first generation dominates; (II.A) For intermediate gate error probability,
but poor coupling efficiency or slow local operation {[}$0.1\frac{L_{att}}{L_{tot}}\apprle\epsilon_{G}\apprle1\%$
and ($\eta_{c}\apprle90\%$ or $t_{0}\apprge1\mu s$){]}, the second
generation \textit{with} encoding is more favorable; (II.B) For
low gate error probability, but low coupling efficiency or slow local
operation {[}$\epsilon_{G}\apprle0.1\frac{L_{att}}{L_{tot}}$ and
($\eta_{c}\apprle90\%$ or $t_{0}\apprge1\mu s$){]}, the second generation
\textit{without} encoding is more favorable; (III) For high coupling
efficiency, fast local operation, and low gate error probability ($\eta_{c}\apprge90\%$,
$t_{0}\apprle1\mu s$, $\epsilon_{G}\apprle1\%$), the third generation
becomes the most favorable scheme in terms of the cost coefficient.

\section{Discussions}

So far, we have mostly focused on the standard proceDure of HEG and
HEP \cite{Briegel1998,Sangouard2011,Deutsch96,Dur99}, the CSS-type
quantum error correcting codes, and the teleportation-based QEC,
which all can be improved and generalized. We have also assumed the
simple cost function that scales linearly with the communication time
and the total number of qubits. In practice, however, the cost function
may have a more complicated dependence on various resources. Nevertheless,
we may extend our analysis by using more realistic cost functions
to compare various QR protocols. As we bridge the architectural
design of QRs and the physical implementations, we may include more
variations of HEG, HEP as well as QEC (such as all optical schemes \cite{Ewert2015,Munro2015}) and use more realistic cost functions,
while the general trend and different parameter regions should remain
mostly insensitive to these details. 

The classification of QR protocols with different performance in the
parameter space also provides a guideline for optimized architectural
design of QRs based on technological capabilities, which are closely
related to physical implementations, including atomic ensembles, trapped
ions, NV centers, quantum dots, nanophotonic devices, etc. (1) The
atomic ensemble can be used as quantum memory with high coupling efficiency
($>80\%$ \cite{Hosseini11,ChenYH13}) and compatible with HEG for the first generation of QRs  \cite{Duan2001}.
An important challenge for ensemble-based QRs is the use of non-deterministic
quantum gates, which can be partly compensated by \textit{multiplexing}
various internal modes of the ensemble memory \cite{Sangouard2011,Bonarota11}.
Alternatively, the atomic ensemble approach can be supplemented by deterministic atom-photon and atom-atom gates using Rydberg blockade, which   can dramatically improve the performance of atomic ensemble approaches and make them compatible with both first and second generations of QRs  \cite{Lukin01,Peyronel12}. (2) The trapped ions, NV centers, and quantum dots all can implement
local quantum operations deterministically \cite{Chiaverini04,NiggD14,sciencekim,Waldherr14,Medford13},
as well as HEG \cite{Moehring2007,Togan10,Bernien12,DeGreve12}. In
principle, they are all compatible with the first and second generations
of QRs. Although the coupling efficiency is relatively low for single
emitters compared to ensembles, it can be boosted with
cavity Purcell enhancement \cite{Kim2011} (by two orders of magnitude). With high
coupling efficiency \cite{Casabone2015a, Meyer2015}, these systems can also be used for
the third generation of QRs. (3) The system of nanophotonic cavity
with individual trapped neutral atoms has recently demonstrated quantum
optical switch controlled by a single atom with high coupling efficiency \cite{Tiecke14,Shomroni14},
which can be used for deterministic local encoding and QEC for the
third generation of QRs. Realization of similar techniques with atom-like emitters are likewise being explored. (4) The opto-electro-mechanical systems have
recently demonstrated efficient coherent frequency conversion between
optical and microwave photons \cite{Bagci14,Andrews14} and can potentially
enable using superconducting systems \cite{Devoret13} for reliable
fast local quantum gates for QRs. 

\section{Conclusion}

In this work, we have classified various QR protocols into
three generations based on different methods for suppressing loss and
operation errors. Introducing the cost function to characterize both
temporal and physical resources, we have systematically compared three
generations of QRs for various experimental parameters, including
coupling efficiency, gate fidelity, and gate times. There are different
parameter regions with drastically different architectural designs
of quantum repeaters with different possible physical implementations. Our work will provide a guideline for the optimal design of quantum networks and help in the extension of quantum network of clocks \cite{Komar2014}, interferometric telescopes \cite{Gottesman2012} and distributed quantum computation \cite{JTSL07b, Monroe2014a} to global scales. In the future, the integration of different generations of QRs will enable the creation of a secure quantum internet \cite{Kimble2008}.
 
\section*{Acknowledgements}

This work was supported by the DARPA Quiness program, ARL CDQI program, ARO, AFOSR, NBRPC (973 program), the Alfred P. Sloan Foundation and the Packard Foundation. We thank Anna Wang, Hong Tang, Ryo Namiki, Prasanta Panigrahi and Steven Girvin for discussions. 

\bibliographystyle{apsrev4-1}
\bibliography{ref.bib}

\newpage
\onecolumngrid

\begin{centering}
\section*{\LARGE \textbf{Supplementary Material}}
\end{centering}
\section*{\large Descriptions of error models}

\subsection*{a) Two-qubit gate error}

Local two-qubit gates, e.g. CNOT gate, are characterized by the gate
infidelity $\epsilon_{G}$. With probability $1-\epsilon_{G}$ the
desired two-qubit gate is applied, while with probability $\epsilon_{G}$
the state of the two qubits becomes a maximally mixed state. Mathematically
the imperfect two-qubit operation on qubit i and j can be expressed
as

\begin{equation}
U\rho U^{\dagger}=(1-\epsilon_{G})U_{ij}\rho U_{ij}^{\dagger}+\frac{\epsilon_{G}}{4}Tr_{ij}[\rho]\otimes I_{ij},
\end{equation}

where $U_{ij}$ stands for perfect two-qubit operation on qubit i
and j, $Tr_{ij}[\rho]$ the partial trace over qubit i and j, and
$I_{ij}$ the identity operator for qubits i and j.

\subsection*{b) Measurement error}

Qubit measurement error is described by the measurement infidelity
$\text{\ensuremath{\xi}}$, which is the probability of a wrong measurement.
The error models for projective measurements of states $|0\rangle$and
$|1\rangle$ are
\begin{eqnarray}
P_{0} & = & (1-\xi)|0\rangle\langle0|+\xi|1\rangle\langle1|\nonumber \\
P_{1} & = & (1-\xi)|1\rangle\langle1|+\xi|0\rangle\langle0|.
\end{eqnarray}

The measurement error can be suppressed by introducing an ancillary
qubit for measurement and measuring both the data and the ancillary
qubits. If the measurement outcomes don't match, it can be considered
as a loss error on that qubit. The contribution of the measurement
error to the overall loss error is negligible given the range of the
gate error rates $(10^{-4}-10^{-2})$ we are considering; if they
match, then the effective measurement error is given by $\frac{\epsilon_{G}}{4}$\cite{key-7-1}.

\subsection*{c) Memory life-times}

In the calculations, the memory qubits are assumed to be perfect,
i.e. their life-time is tremendously longer than any characteristic
times involved in each scheme. In this sense the most demanding scheme
is the first generation, which is optimum at long communication distances,
e.g. $L_{tot}=10^{4}km$, and high gate errors. The required coherence
time $\tau$ for memory qubits is at least limited by the fundamental
two-way classical communication time between Alice and Bob 
\begin{equation}
t_{c1}=\frac{L_{tot}}{c}\sim50ms,
\end{equation}
where $c=2\times10^{5}km/s$ is the speed of light in telecom-wavelength
optical fiber. Recent experiments with trapped ions, superconducting
qubits, solid state spins and neutral atoms have demonstrated quantum
memory life-times approaching or exceeding this characteristic value.
For the second generation, the characteristic communication time is
\begin{equation}
t_{c2}=\frac{L_{att}}{c}\sim100\mu s
\end{equation}
where $L_{att}=20$km at telecomm wavelength. The corresponding coherence
times are far less demanding than that of the first generation, which
relieves the strong life-time requirements on memory qubits and makes
the two generations more plausible in practice. Note that when the
operation time $t_{0}$ becomes comparable or larger than the characteristic
communication time, it is then the operation time $t_{0}$ that puts
limits to the coherence time $\tau$. Third generation QRs are not
limited by the two way communication time because it is a fully one
way communication scheme.

\section*{Quantum entanglement generation, purification and connection}

\subsection*{a) Generation of elementary entangled pairs}

\subsubsection*{Heralded entanglement generation with two-photon detection}

Using two photon-detection in the middle\cite{key-1,key-8}, the success
probability of one trial of generating entanglement between two memory
qubits in neighboring stations is 

\begin{equation}
p=\frac{1}{2}\eta_{c}^{2}e^{-L_{0}/L_{att}},
\end{equation}
where $L_{0}$ is the spacing between neighboring stations and $\eta_{c}$
is the coupling efficiency accounting for the emission of the photon
from the memory qubit, ``upload'' of the photon into the optical
fiber, ``download'' of the photon from the fiber and the final detection
of photons.

\subsubsection*{Fidelity}

Practically an entangled pair generated between neighboring stations
may not be a perfect Bell pair and its state is characterized by a
density matrix
\begin{equation}
\rho=a|\varphi^{+}\rangle\langle\varphi^{+}|+b|\varphi^{-}\rangle\langle\varphi^{-}|+c|\psi^{+}\rangle\langle\psi^{+}|+d|\psi^{-}\rangle\langle\psi^{-}|,
\end{equation}
where $|\varphi^{\pm}\rangle=\frac{1}{\sqrt{2}}(|00\rangle\pm|11\rangle)$
and $|\psi^{\pm}\rangle=\frac{1}{\sqrt{2}}(|01\rangle\pm|10\rangle)$
are the four Bell states. The fidelity of the pair is thus defined
as 
\begin{equation}
F\equiv a=\langle\varphi^{+}|\rho|\varphi^{+}\rangle,
\end{equation}

Both the first and second generations QRs rely on generating elementary
entangled pairs between neighboring repeater stations and then extending
the entanglement to longer distances. With the technique of HEP (see
the next section), the fidelity of entangled pairs can be boosted
to near-unity at the cost of reducing the total number of them and
purified pairs can be connected to obtain longer entangled pairs or
used as resources for the implementation of remote quantum gates.

However, with imperfect quantum operations and measurements, there
is an upper bound on the fidelity of entangled pairs even with entanglement
purification. It is in general, a function of the density matrix of
raw Bell pairs $\rho$, gate infidelity $\epsilon_{G}$ and measurement
infidelity $\xi$, and depends on the specific purification protocol
one uses. Using Deutsch purification protocol (see the next section),
the value of this upper bound can be approximated as 
\begin{eqnarray}
F_{u.b.} & = & 1-\frac{5}{4}\epsilon_{G}-(\frac{9}{4}\xi+\frac{19}{4}\epsilon_{G})\epsilon_{G}+\mathcal{O}(\epsilon_{G,}\xi)^{3}\nonumber \\
 & \approx & 1-\frac{5}{4}\epsilon_{G},
\end{eqnarray}
in which we assume depolarized states for input raw Bell pairs. This
approximate expression holds at small $\epsilon_{G}'s$ $(\apprle1\%)$.
In our calculations and comparison, the temporal resources and physical
resources consumed in obtaining purified pairs at the elementary level
are not accounted; elementary entangled pairs generated between neighboring
stations are assumed to directly take this asymptotic value. The associated
additional cost in the purification can be easily added as an overhead
into the cost function.

\subsection*{b) Deutsch and D\"ur purification protocols}

In general, despite differences in experimental requirements and efficiencies,
the choice of purification protocols will not change the big picture.
In this paper, we mainly consider two widely used entanglement purification
protocols: Deutsch protocol\cite{key-7} and D\"ur\cite{key-9} protocol.
Compared to other purification schemes, the Deutsch protocol reaches
higher fidelities with fewer rounds of purification so its upper bound
is used as the fidelity of elementary pairs between neighboring stations.
The D\"ur purification protocol is very similar to the Deutsch protocol,
except that one of the two pairs, call auxiliary pair,  is never discarded
and will be prepared in the same state in each round of purification.
This is sometimes also call ``entanglement pumping''. The D\"ur
purification protocol saves qubit resources by keeping making use
of the auxiliary pair, while the state preparation in each purification
round is costly in time as a trade-off and if one round fails, the
whole purification needs to be started over again.

Here we study the purification with two input pairs characterized
by density matrices $\rho_{1}$ and $\rho_{2}$ ($\rho_{1}=\rho_{2}$
in the case of Deutsch protocol). As mentioned in the previous section,
we express the density matrices in the Bell basis \{$|\varphi^{+}\rangle$,
$|\varphi^{-}\rangle$, $|\psi^{+}\rangle$, $|\psi^{-}\rangle$\}.
With input states\{$a_{1},b_{1},c_{1},d_{1}$\} and \{$a_{2},b_{2},c_{2},d_{2}$\},
in the presence of gate infidelity $\epsilon_{G}$ and measurement
infidelity $\xi$, the success probability $P$ and the purified state
characterized by the diagonal elements \{$a,b,c,d$\} are the following
\begin{eqnarray}
P & = & (1-\epsilon_{G})^{2}\{[\xi^{2}+(1-\xi)^{2}][(a_{1}+d_{1})(a_{2}+d_{2})+(b_{1}+c_{1})(c_{2}+b_{2})]+2\xi(1-\xi)[(a_{1}+d_{1})(b_{2}+c_{2})+(b_{1}+c_{1})(a_{2}+d_{2})]\}\nonumber \\
 &  & +\frac{1}{2}[1-(1-\epsilon_{G})^{2}]\nonumber \\
a & = & \frac{1}{P}\{(1-\epsilon_{G})^{2}[(\xi^{2}+(1-\xi)^{2}](a_{1}a_{2}+d_{1}d_{2})+2\xi(1-\xi)(a_{1}c_{2}+d_{1}b_{2})\}+\frac{1}{8}\text{[}1-(1-\epsilon_{G})^{2})]\nonumber \\
b & = & \frac{1}{P}\{(1-\epsilon_{G})^{2}[(\xi^{2}+(1-\xi)^{2}](a_{1}d_{2}+d_{1}a_{2})+2\xi(1-\xi)(a_{1}b_{2}+d_{1}c_{2})\}+\frac{1}{8}\text{[}1-(1-\epsilon_{G})^{2})]\nonumber \\
c & = & \frac{1}{P}\{(1-\epsilon_{G})^{2}[(\xi^{2}+(1-\xi)^{2}](b_{1}b_{2}+c_{1}c_{2})+2\xi(1-\xi)(b_{1}d_{2}+c_{1}a_{2})\}+\frac{1}{8}\text{[}1-(1-\epsilon_{G})^{2})]\nonumber \\
d & = & \frac{1}{P}\{(1-\epsilon_{G})^{2}[(\xi^{2}+(1-\xi)^{2}](b_{1}c_{2}+c_{1}b_{2})+2\xi(1-\xi)(b_{1}a_{2}+c_{1}d_{2})\}+\frac{1}{8}\text{[}1-(1-\epsilon_{G})^{2})]
\end{eqnarray}

\subsection*{c) Entanglement Swapping}

Entanglement swapping is used in the first generation and second generation
without encoding to extend the distance of entanglement. With imperfect
CNOT operation and measurements, the diagonal elements \{$a,b,c,d$\}
in the Bell basis of the resulting state obtained from connecting
deterministically the input pairs \{$a_{1},b_{1},c_{1},d_{1}$\} and
\{$a_{2},b_{2},c_{2},d_{2}$\} is

\begin{eqnarray}
a & = & (1-\epsilon_{G})\{(1-\xi)^{2}(a_{1}a_{2}+b_{1}b_{2}+c_{1}c_{2}+d_{1}d_{2})+\xi(1-\xi)[(a_{1}+d_{1})(b_{2}+c_{2})+(b_{1}+c_{1})(a_{2}+d_{2})]\nonumber \\
 &  & +\text{\ensuremath{\xi}}^{2}(a_{1}d_{2}+d_{1}a_{2}+b_{1}c_{2}+c_{1}b_{2})\}+\frac{\epsilon_{G}}{4}\nonumber \\
b & = & (1-\epsilon_{G})\{(1-\xi)^{2}(a_{1}b_{2}+b_{1}a_{2}+c_{1}d_{2}+d_{1}c_{2})+\xi(1-\xi)[(a_{1}+d_{1})(a_{2}+d_{2})+(b_{1}+c_{1})(b_{2}+c_{2})]\nonumber \\
 &  & +\text{\ensuremath{\xi}}^{2}(a_{1}c_{2}+c_{1}a_{2}+b_{1}d_{2}+d_{1}b_{2})\}+\frac{\epsilon_{G}}{4}\nonumber \\
c & = & (1-\epsilon_{G})\{(1-\xi)^{2}(a_{1}c_{2}+c_{1}a_{2}+b_{1}d_{2}+d_{1}b_{2})+\xi(1-\xi)[(a_{1}+d_{1})(a_{2}+d_{2})+(b_{1}+c_{1})(b_{2}+c_{2})]\nonumber \\
 &  & +\text{\ensuremath{\xi}}^{2}(a_{1}b_{2}+b_{1}a_{2}+c_{1}d_{2}+d_{1}c_{2})\}+\frac{\epsilon_{G}}{4}\nonumber \\
d & = & (1-\epsilon_{G})\{(1-\xi)^{2}(a_{1}d_{2}+d_{1}a_{2}+c_{1}b_{2}+b_{1}c_{2})+\xi(1-\xi)[(a_{1}+d_{1})(b_{2}+c_{2})+(b_{1}+c_{1})(a_{2}+d_{2})]\nonumber \\
 &  & +\text{\ensuremath{\xi}}^{2}(a_{1}a_{2}+b_{1}b_{2}+c_{1}c_{2}+d_{1}d_{2})\}+\frac{\epsilon_{G}}{4}
\end{eqnarray}

Note that deterministic entanglement swapping is crucial for long
distance quantum communication using QRs. Otherwise the success probability
of entangling two qubits separated by $L_{tot}$ drops exponentially
as $L_{tot}$ increases.

\section*{Implementation and optimization}

\subsection*{a) First generation}

The first generation of QRs corrects photon loss and operation errors
with HEG and HEP, respectively. To overcome the exponential decay
in key generation rate induced by photon loss, the total distance
$L_{tot}$ is divided into $2^{n}$ segments (n is called nesting
level\cite{key-1,key-9}) and elementary entangled pairs are generated
within each segment, i.e. over repeater spacing $L_{0}=\frac{L_{tot}}{2^{n}}$.
An entangled pair covering the total distance $L_{tot}$ can be generated
via n levels of entanglement swapping: at each level, two adjacent
entangled pairs are connected so that an entangled pair over twice
the distance is produced.

However, entanglement swapping necessarily reduces the fidelity of
the entangled pairs due to the following two reasons: 1) entanglement
swapping involves CNOT operation and measurements, which themselves
are imperfect in reality and will introduce noise. 2) Despite imperfect
operations, the connection of two imperfect Bell pairs gives a pair
with lower fidelity. So multiple rounds of entanglement purification
may need to be incorporated at each level to maintain the fidelity,
so that the final pair covering $L_{tot}$ is sufficiently robust
for secure key distribution. 

In the optimization, to determine the best scheme from the first generation
QRs, we first fix
\begin{itemize}
\item Total distance $L_{tot}$,
\item Coupling efficiency $\eta_{c}$ 
\item Gate error rate $\epsilon_{G}$ , and thus the fidelity of elementary
Bell pairs $F_{0}=1-\frac{5}{4}\epsilon_{G}$
\item Gate time $t_{0}$
\end{itemize}
and we vary the following parameters:
\begin{itemize}
\item Number of nesting level: $N$
\item Number of rounds of purification at each level: $\overrightarrow{M}=(M_{0},M_{1},M_{2},\cdots,M_{N})$
\item Choice of entanglement purification protocol: Deutsch or D\"ur
\end{itemize}
In carrying out the time resource consumed in generating one remote
pair, we adopted similar approximations and derivations as in a previous
work\cite{key-1}, with three major changes:
\begin{enumerate}
\item We allow arbitrary number of rounds of purification at each level
in the optimization, and hence schemes selected could potentially
be better optimized.
\item Without losing generality, we use four-state protocol, instead of
six-state protocol, in calculating the secure fraction at the asymptotic
limit.
\item For long-distance quantum communication, we only consider \textit{deterministic}
entanglement swapping and take into account the gate operation time
$t_{0}$.
\end{enumerate}
For a given set of $N$ and $\overrightarrow{M}$, detailed derivations
of expressions of temporal and physical resource consumed in Deutsch
and D\"ur are given below.

\subsubsection*{Deutsch et al. entanglement purification protocol}

The temporal resource to generate one bit of raw key, $T_{Deu}$,
can be calculated as follows:

\begin{eqnarray}
T_{Deu} & =T_{0}\cdot\{ & (\frac{3}{2})^{N}({\prod_{x=0}^{N-1}}\text{\ensuremath{A_{Deu}}[N-x]})(\frac{1}{P_{0}}A_{Deu}[0]+B_{Deu}[0])+{\sum_{y=1}^{N}}(\frac{3}{2})^{N-y}B_{Deu}[y]}{\prod_{x=0}^{N-(y+1)}\text{\text{\ensuremath{A_{Deu}}[N-x]}}\nonumber \\
 &  & +\frac{t_{0}}{T_{0}}{\sum_{y=1}^{N}}(\frac{3}{2})^{N-y}{\prod_{x=0}^{N-y}}\text{\text{\ensuremath{A_{Deu}}[N-x]\}},}
\end{eqnarray}

where

\begin{eqnarray}
A_{Deu}[i] & \equiv & (\frac{3}{2})^{M_{i}}{\prod_{x=0}^{M_{i}-1}}\frac{1}{P_{Deu}(M_{i}-x,i)}\nonumber \\
B_{Deu}[i] & \equiv & (\frac{t_{0}}{T_{0}}+2^{i}){\sum_{y=0}^{M_{i}-1}}(\frac{3}{2})^{y}{\prod_{x=0}^{y}}\frac{1}{P_{Deu}(M_{i}-x,i)}.
\end{eqnarray}
Here, $A_{Deu}[i]$ accounts for the time consumed in the preparation
of Bell pairs for purification, and $B_{Deu}[i]$ includes gate operation
time and the \textit{two-way} classical signaling time associated
with confirming the success of purification. $P_{Deu}(i,j)$ is the
success probability of the $i^{th}$-round of purification at the
$j^{th}$ nesting level with Deustch purification protocol. $T_{0}=\frac{L_{0}}{c}$
is the time unit for the \textit{two-way} classical signaling between
neighboring stations, where $c=2\times10^{5}km/s$ in optical fiber.
The secure key generation rate, $R_{secure}$ (sbit/s), can be written
as
\begin{equation}
R_{secure}^{Deu}=r_{secure}\cdot\frac{1}{T_{Deu}},
\end{equation}
where $r_{secure}$ is the asymptotic secure fraction and in the four-state
protocol can be approximately expressed as 
\begin{equation}
r_{secure}=Max[1-2h(Q),0],
\end{equation}
where $Q=\frac{Q_{X}+Q_{Z}}{2}$ is the average quantum bit error
rate (QBER) and $h(Q)=-Q\log_{2}Q-(1-Q)\log_{2}(1-Q)$ is the binary
entropy function. $Q_{X/Z}$ can be calculated from the density matrix
of the entangled shared by Alice and Bob in the end. The physical
resource, in terms of the number of memory qubits, consumed at half
a station can be written as
\begin{equation}
Z_{Deu}=2^{{\sum_{i=0}^{N+1}}M_{i}},
\end{equation}
and the cost function becomes
\begin{equation}
C=\text{\ensuremath{\frac{2^{N+1}\cdot Z_{Deu}}{R_{secure}^{Deu}}}.}
\end{equation}

\subsubsection*{D\"ur et al. entanglement purification protocol}

The temporal resource to generate one bit of raw key, $T_{\text{D\"ur}}$,
can be calculated as follows:

\begin{eqnarray}
T_{\text{D\"ur}} & =T_{0}\cdot\{ & (\frac{3}{2})^{N}(\frac{1}{P_{0}}A_{\text{D\"ur}}[0]+B_{\text{D\"ur}}[0])(}{\prod_{x=0}^{N-1}\text{\ensuremath{A_{D\ddot{u}r}}[N-x]})+{\sum_{y=1}^{N}}(\frac{3}{2})^{N-y}B_{\text{D\"ur}}[y]{\prod_{x=0}^{N-(y+1)}}\text{\text{\ensuremath{A_{D\ddot{u}r}}[N-x]}}\nonumber \\
 &  & +\frac{t_{0}}{T_{0}}{\sum_{y=1}^{N}}(\frac{3}{2})^{N-y}{\prod_{x=0}^{N-y}}\text{\text{\ensuremath{A_{D\ddot{u}r}}[N-x]\}},}
\end{eqnarray}
where,
\begin{eqnarray}
A_{\text{D\"ur}}[i] & \equiv & {\prod_{x=0}^{M_{i}-1}}\frac{1}{P_{\text{D\"ur}}(M_{i}-x,i)}+\sum_{y=0}^{M_i-1}{\prod_{x=0}^{y}}\frac{1}{P_{\text{D\"ur}}(M_{i}-x,i)}\nonumber \\
B_{\text{D\"ur}}[i] & \equiv & (\frac{t_{0}}{T_{0}}+2^{i}){\sum_{y=0}^{M_{i}-1}}{\prod_{x=0}^{y}}\frac{1}{P_{\text{D\"ur}}(M_{i}-x,i)}.
\end{eqnarray}
Here, $A_{\text{D\"ur}}[i]$ accounts for the time consumed in the preparation
of Bell pairs for purification (notice the extra term due to entanglement
pumping), and $B_{\text{D\"ur}}[i]$ includes gate operation time and the
\textit{two-way} classical signaling time associated with confirming
the success of purification. $P_{\text{D\"ur}}(i,j)$ is the success probability
of the $i^{th}$-round of purification at the $j^{th}$ nesting level
with D\"ur purification protocol. Because of the unique entanglement
pumping mechanism, the physical resource consumed at half a station
is reduced compared to Deustch purification protocol and expressed
as
\begin{equation}
Z_{\text{D\"ur}}=N+2-|\{M_{i}:M_{i}=0\}|.
\end{equation}
The derivations of secure key generation rate and hence cost function
are similar to those in the previous section.

\subsection*{b) Second generation without encoding}

The second generation of QRs relies on generating encoded Bell pairs
between neighboring stations and performing error correction during
entanglement swapping at the encoded level. With encoding and error
correction, physical gate error rates and imperfections in raw Bell
pairs are suppressed to higher orders and thus entanglement can be
extended to very long distances with high fidelity. However, if we
are interested in the best schemes at low gate error rate $\epsilon\apprle10^{-3}$
and total distance $L_{tot}\sim10^{3}km$, the encoding may turn out
unnecessary and resources can be saved by simply generating elementary
pairs between neighboring stations and implementing entanglement swapping.
Fixing the same parameters as the previous section, we vary the following
parameters
\begin{itemize}
\item Number of memory qubits per half station: $N$
\item Spacing between neighboring stations: $L_{0}$
\item Number of rounds of elementary entanglement generation trial: $n_{E.G.}$
\end{itemize}
The secure key generation rate, $R_{secure}$ (sbit/s), can be written
as 
\begin{equation}
R_{secure}=\frac{[1-Prob(0,n_{E.G.})]^{\lceil\frac{L_{tot}}{L_{0}}\rceil}\cdot r_{secure}}{n_{E.G.}\cdot(T_{0}+t_{0})},
\end{equation}
where basic communication time $T_{0}=\frac{L_{0}}{c}$ and $r_{secure}$
follows the same definition above. $Prob(i,n_{0})=\left(\begin{array}{c}
M\\
n_{0}
\end{array}\right)p^{n_{0}}(1-p)^{M-n_{0}}$ is the probability to generate $i$ elementary pairs with $M$ qubits
in the two half nodes after $n_{0}$ rounds of entanglement generation.
Therefore, $1-Prob(0,n_{E.G.})$ means the probability to have \emph{at least}
one entangled pair between two neighboring stations. Note that frequency
and spatial multiplexing may needed to be incorporated during the
transmission of flying qubits and entanglement swapping, respectively.
The cost function that will be optimized can be written as 
\begin{equation}
C=\frac{2M\cdot\lceil\frac{L_{tot}}{L_{0}}\rceil}{R_{secure}}.
\end{equation}

\subsection*{c) Second generation with encoding}

For second generation QRs with encoding, encoded Bell pairs are created
between neighboring repeater stations and later an encoded entanglement
swapping operation is performed at every QR station to generate an
encoded Bell pair between distant stations. As in the case of first
generation QRs, Bell pairs are generated using HEG between neighboring
stations with a high fidelity. These Bell pairs are used as a resource
to perform teleportation based CNOT gates between neighboring stations,
thereby realizing an encoded CNOT operation between neighboring QR
stations. The depolarization error on the data qubits can be modeled
as

\begin{equation}
\rho'=\mathcal{E}\left(\rho\right)=\left(1-\epsilon_{d}\right)\rho+\frac{\epsilon_{d}}{4}\sum_{k=0}^{3}\sigma_{k}\rho\sigma_{k}.
\end{equation}
The probability of an error being detected in any one (X or Z) of
the measurements is given by, 

\begin{equation}
\epsilon_{X/Z}=\epsilon_{d}+\epsilon_{G}+2\xi+\frac{2}{3}(1-F_{0})+O(\epsilon_{G},\xi)^{2}
\end{equation}
Any $[[N,k,2t+1]]$ CSS code can correct up to $t$ X-errors and $t$
Z-errors respectively. Taking this into account, the probability of
correctly and incorrectly decoding the qubit are given by, 

\begin{equation}
p_{correct(X/Z)}^{2G}=\sum_{k=0}^{t}\left(\begin{array}{c}
N\\
k
\end{array}\right)\epsilon_{X/Z}^{k}(1-\epsilon_{X/Z})^{(N-k)},
\end{equation}
\begin{equation}
p_{incorrect(X/Z)}^{2G}=\sum_{k=t+1}^{N}\left(\begin{array}{c}
N\\
k
\end{array}\right)\epsilon_{X/Z}^{k}(1-\epsilon_{X/Z})^{(N-k)}
\end{equation}
respectively. Accounting for logical errors in odd number of repeater
stations, quantum bit error rates for X and Z basis after R repeater
stations is given by, 

\begin{equation}
Q_{(X/Z)}=\frac{1}{2}\left[1-\left(p_{correct(X/Z)}^{2G}-p_{incorrect(X/Z)}^{2G}\right)^{R}\right].
\end{equation}
Where the effective quantum bit error rate is given by$Q=\frac{1}{2}\left(Q_{X}+Q_{Z}\right).$
The success probability of the protocol is conditioned on having enough
Bell pairs between neighboring stations to apply a teleportation based
CNOT gate. We can then obtain the key generation rates similar to
the case of second generation without encoding. We consider the Steane
{[}{[}7,1,3{]}{]} code, Golay {[}{[}23,1,7{]}{]} code and the QR {[}{[}103,1,19{]}{]}
codes in our optimization.

\subsection*{d) Third generation QRs}

Third generation QRs rely on encoded qubits to relay data from one
repeater station to the next where an error correction operation is
performed. Since there is just one round of upload and download between
the memory qubit and the fiber for third generation QRs, the probability
that the photon reaches the neighboring station is given by $\eta_{c}e^{-L_{0}/L_{att}}$.
Unlike second generation QRs with encoding, teleportation based error
correction (TEC) is performed within every repeater station locally
for third generation QRs. Teleportation based error correction requires
an encoded CNOT gate between the incoming encoded qubit block (with
loss and operation errors) and encoded qubit block (with no loss errors)
at every repeater station. R is the incoming encoded block and S is
the encoded block from the encoded Bell pair. The depolarization errors
on blocks R and S can be modeled as,

\begin{equation}
\rho_{R}^{\prime}=\mathcal{E}_{R}\left(\rho_{RS}\right)=\eta\left(1-\epsilon_{d}\right)\rho_{R}+\frac{\eta\epsilon_{d}}{4}\sum_{k=0}^{3}\sigma_{k}\rho\sigma_{k}+\left(1-\eta\right)|vac\rangle\langle vac|
\end{equation}

\begin{equation}
\rho'_{S}=\mathcal{E}_{S}\left(\rho_{RS}\right)=\left(1-\epsilon_{d}\right)\rho_{S}+\frac{\epsilon_{d}}{4}\sum_{k=0}^{3}\sigma_{k}\rho\sigma_{k},
\end{equation}
To be consistent with\cite{key-6}, we make the following assumptions
in our analysis. 1) Errors are not propagated between repeater stations.
2) Each qubit has an independent error. The effective X/Z error detected
at the measurement (Y errors are detected in both X and Z measurements)
is given by, 

\begin{equation}
\epsilon_{X/Z}=(\epsilon_{d}+\frac{\epsilon_{G}}{2}+\xi)\eta+O(\epsilon_{G,},\xi)^{2}.
\end{equation}
We consider (n,m) quantum parity codes given by

\begin{equation}
|\pm\rangle_{L}=\frac{1}{2^{n/2}}(|0\rangle^{\otimes m}\pm|1\rangle^{\otimes m})^{\otimes n},
\end{equation}
in our analyses. The outcome of the measurement of the logical operators
$X_{L}$ and $Z_{L}$ for TEC can be determined through a majority
voting procedure discussed in detail in \cite{key-6}. There are three
possible outcomes of the majority voting procedure: a) Heralded failure
leading to the inability to perform a majority voting with probability
$p_{unknown(X/Z)}^{3G}$. b) Perform a majority voting and correctly
decoding the qubit with probability $p_{correct(X/Z)}^{3G}$. c) Perform
a majority voting and incorrectly decoding the qubit with probability
$p_{incorrect(X/Z)}^{3G}$. Here, we treat the three events as independent
for the measurement of the logical operators for simplicity. The success
probability accounting for no heralded failure in any one of the $R$
repeater stations is given by, 

\begin{equation}
P_{succ}=(1-p_{unknown(X/Z)}^{3G})^{R}
\end{equation}
Accounting for errors in odd number of repeater stations, the quantum
bit error rate for $X/Z$ bases is given by, 

\begin{equation}
Q_{(X/Z)}=\frac{1}{2}\left[1-\left(\frac{p_{correct(X/Z)}^{3G}-p_{incorrect(X/Z)}^{3G}}{p_{correct(X/Z)}^{3G}+p_{incorrect(X/Z)}^{3G}}\right)^{R}\right]
\end{equation}
The asymptotic secure key generation rates is given by, 

\begin{equation}
R_{secure}=Max[\frac{P_{succ}}{t_{0}}\{1-2h(Q)\},0],
\end{equation}
where $t_{0}$ is the time taken to apply local operations. The cost
function for the $(n,m)$ quantum parity codes is given by

\begin{equation}
C=\frac{2m\cdot n\lceil\frac{L_{tot}}{L_{0}}\rceil}{R_{secure}}.
\end{equation}
For a fair comparison with second generation QRs, where a largest
code of $[[103,1,19]]$ code was used, we restrict the maximum qubits
for the third generation quantum repeaters to be 200. We restrict
the search of $(n,m)$ quantum parity codes within the range $2\leq(m,n)\leq20$.


\end{document}